\documentstyle[preprint,tighten,aps]{revtex}
\begin{document}
\preprint{\vbox{ \null\hfill DFTT 18/97 \\
\null\hfill INFNCA-TH9703 \\
\null\hfill hep-ph/9703303 \\
\vspace{1.0truecm}}}
\draft
\newcommand{\beq}{\begin{equation}}
\newcommand{\eeq}{\end{equation}}
\newcommand{\barr}{\begin{eqnarray}}
\newcommand{\earr}{\end{eqnarray}}
\newcommand{\pup}{p^\uparrow}
\newcommand{\pdown}{p^\downarrow}
\newcommand{\la}{\lambda}
\def\nostrocostruttino#1\over#2{\mathrel{\mathop{\kern 0pt \rlap
{\hbox{$#1$}}} \hbox{\kern-.135em $#2$}}}
\def\sumint{\nostrocostruttino \sum \over {\displaystyle\int}}
\newcommand{\bfk}{\mbox{\boldmath $k$}}
\newcommand{\bfy}{\mbox{\boldmath $y$}}
\title{Single-Spin Asymmetries in $\protect\bbox{\pup p \to \pi X}$
 and Chiral Symmetry\\}

\author{ Mauro~Anselmino$^{a}$, Alessandro~Drago$^{b}$, and
 Francesco Murgia$^{c}$}

\address{$^{a}$Dipartimento di Fisica Teorica, Universit\`a di Torino\\
and INFN, Sezione di Torino, V. Giuria 1, 10125 Torino, Italy \\
$^{b}$Dipartimento di Fisica, Universit{\`a} di Ferrara \\
and INFN, Sezione di Ferrara, 44100 Ferrara, Italy \\ 
$^{c}$INFN, Sezione di Cagliari, Via A. Negri 18, 09127 Cagliari, Italy 
\\}
\date{March 1997}
\maketitle
\begin{abstract}
We study a possible origin of single spin asymmetries in the large $p_T$ 
inclusive production of hadrons in the scattering of unpolarized protons
or leptons on transversely polarized nucleons. Such asymmetry is related 
to the single-spin asymmetry for the process $\pup \to q+X$ or, 
equivalently, to the off-diagonal matrix element of the quark density 
operator $\bar \psi \gamma^+ \psi$; this quantity need not be zero if 
transverse momentum and spin-isospin interactions are taken into account.
The different signs for the large single spin asymmetries observed
in $\pup p \to \pi^+ X$ and $\pup p \to \pi^- X$ can be explained as a 
consequence of chiral symmetry. Crucial tests of the model are suggested.
\end{abstract}
\pacs{PACS Number(s): 13.88.+e, 13.85.Ni, 12.38.-t, 12.39.Fe}

\narrowtext

It has become very popular to use chiral Lagrangians to study nucleon 
spin structure functions. They have been used, for example, to understand
the reduction of the proton spin fraction carried by the 
quarks \cite{chengli}. In this paper we show that chiral symmetry 
plays a crucial r\^ole in the single-spin asymmetries observed in the 
inclusive production of pions in the scattering of transversely 
polarized protons off unpolarized ones \cite{data}. In particular we 
shall show that the opposite sign of the asymmetry for positive and 
negative pions can be related to the underlying chiral symmetry of the 
model describing the nucleon. This mechanism can be tested by measuring
single-spin asymmetries 
in inclusive hadronic processes, $\pup p \to \gamma X$
and $\pup p \to h X$, or in the Deep Inelastic Scattering (DIS) of
unpolarized (or longitudinally polarized) leptons 
on transversely polarized nucleons.

The main point of this paper is the analysis of a new quantity 
introduced in Refs. \cite{sivers} and \cite{mauro}:
\barr
&\phantom{=}&\Delta^Nf_{a/\pup}(x_a,\bfk_{\perp a})\nonumber\\
 &\equiv& \sum_{\la^{\,}_a} \left[
\hat f_{a, \la^{\,}_a / \pup} (x_a,\bfk_{\perp a})
- \hat f_{a, \la^{\,}_a / \pdown} (x_a,\bfk_{\perp a}) \right]
\label{isiv1}\\
&=& \sum_{\la^{\,}_a}
\left[ \hat f_{a, \la^{\,}_a / \pup} (x_a,\bfk_{\perp a})
- \hat f_{a, \la^{\,}_a / \pup} (x_a, -\bfk_{\perp a}) \right] \,,
\label{isiv2}
\earr
where $\hat f_{a, \la^{\,}_a / p^{\uparrow (\downarrow)}}(x_a,\bfk_{\perp a})$
is the number density of partons $a$ with helicity $\la^{\,}_a$,
momentum fraction $x_a$ and intrinsic transverse momentum $\bfk_{\perp a}$
in a transversely polarized proton. Eq. (\ref{isiv2}) follows from
Eq. (\ref{isiv1}) by rotational invariance and explicitely shows that
$\Delta^Nf_{a/\pup}(x,\bfk_{\perp}) = 0$ when $\bfk_{\perp}=0$.

This new quantity can be regarded as a single spin asymmetry or analysing
power for the $\pup \to a + X$ process; if we define the polarized number
densities in terms of {\it distribution amplitudes} as
\beq
\hat f_{a, \la^{\,}_a/ \pup}(x_a, \bfk_{\perp a}) = \sumint_{X,\la_{X}}
|{\cal G}^{a/p}_{\la_{X}, \la^{\,}_a; \uparrow}(x_a,\bfk_{\perp a})|^2
\label{g}
\eeq
then we have, in the helicity basis,
\barr
&\phantom{=}&\Delta^Nf_{a/\pup}(x_a,\bfk_{\perp a})\nonumber\\ &=&
2\,\Im{\rm m} \sumint \sum_{\la^{\,}_a} \left[
{\cal G}^{a/p}_{\la_{X}, \la^{\,}_a;+}(x_a,\bfk_{\perp a}) \,\,
{\cal G}^{a/p\,{\textstyle *}}_{\la_{X},\la^{\,}_a;-}(x_a,\bfk_{\perp a}) 
\right] \nonumber \\
&\equiv& 2 \, I^{a/p}_{+-}(x_a,\bfk_{\perp a}) \,,
\label{iap}
\earr
where $+$ and $-$ stand respectively for the proton helicity 
$\la_p = 1/2$ and $\la_p = -1/2$. 
In equations above {\small $\sumint$ } stands for a spin sum and phase space
integration of the undetected particles, considered as a system $X$.
Notice that, in the absence of quark transverse motion ($\bfk_\perp = 0$),
angular momentum conservation implies $\la_p = \la_a + \la_X$ so that,
once more, we see that $I_{+-}(x,\bfk_\perp = 0) = 0$. 
It is useful to compare the previous function with 
well known distribution functions:
\barr
f_1^a(x_a) = \sumint\! \int\! d\bfk_{\perp a}\{
&|&{\cal G}^{a/p}_{\la_{X}, +;+}(x_a,\bfk_{\perp a})|^2+\nonumber\\
&|&{\cal G}^{a/p}_{\la_{X}, -;+}(x_a,\bfk_{\perp a})|^2 \} \\
g_1^a(x_a) = \sumint\! \int\! d\bfk_{\perp a} \{
&|&{\cal G}^{a/p}_{\la_{X}, +;+}(x_a,\bfk_{\perp a})|^2 -\nonumber\\
&|&{\cal G}^{a/p}_{\la_{X}, -;+}(x_a,\bfk_{\perp a})|^2 \} \\
h_1^a(x_a) = \sumint \!\int\! d\bfk_{\perp a} \>
&{\cal G}&^{a/p}_{\la_{X}, +;+}(x_a,\bfk_{\perp a}) \>
{\cal G}^{a/p\,{\textstyle *}}_{\la_{X}, -;-}
(x_a,\bfk_{\perp a})\, .\nonumber\\
&& 
\earr
In the last equation we have neglected terms which vanish when
integrated on $\bfk_\perp$.

According to the usual operatorial definition of quark densities 
using light cone variables \cite{collins} Eq. (4) can be written as 
\barr 
I^{a/p}_{+-}(\!\!&x&\!\!,\bfk_\perp)  \nonumber \\
&=&\Im{\rm m}\int{d y^- d\bfy_\perp\over(2 \pi)^3}
e^{-i x p^+ y^- + i \bfk_\perp\cdot\bfy_\perp}\nonumber\\
&\times&\langle p, -|\bar\psi_a(0,y^-,\bfy_\perp)
{\gamma^+\over 2}\psi_a(0)|p, + \rangle \nonumber \\
&=& \Im{\rm m} \, \hat f_{a/p}(x, \bfk_\perp; -,+) 
\earr 
where in the last line we have adopted the notations of Ref. \cite{collins}.
This definition coincides with the previous one, Eq. (4), as it can be seen 
using completeness: 
$\bar\psi_a\gamma^+\psi_a=\psi^{\dagger}_{a+}\psi_{a+}=
\sum_X\psi^{\dagger}_{a+}|X\rangle\langle X|\psi_{a+}$. The $I_{+-}$ function
corresponds therefore to the imaginary part of the off-diagonal matrix elements
of the same operator appearing in the definition of $f_1$.

In Ref. \cite{collins} it is argued that such off-diagonal matrix elements
are zero due to the time-reversal invariance of QCD, and indeed this is proved 
by exploiting the time-reversal and parity transformation properties of 
free Dirac spinors. In terms of distribution amplitudes for the process
$\pup \to q + X$ this can be understood by observing that no time-reversal
even observable can be constructed with 2 independent momenta and one spin 
vector. In Ref. \cite{mauro} soft initial state interactions between the
colliding protons were invoked to avoid the problem, assuming that such 
interactions do not violate the factorization scheme. This is what happens
in the fragmentation of a transversely polarized quark, which might give 
origin to single spin asymmetries -- the so called Collins or sheared
jet effect \cite{collins,artcol} -- due to final state interactions of
the fragmenting quark. 

We show here that time-reversal invariance does not necessarily imply for the
off-diagonal matrix elements (8) to be zero, even when neglecting initial 
state interactions, provided some spin-isospin interactions are present 
in the quark Lagrangian, as it happens in chiral models. 

In the proof of the vanishing of $I_{+-}$ {\it via} time reversal,
it is crucial that the quark fields $\psi$ are expanded on 
a basis of eigenstates of the Hamiltonian. 
On the other hand, when a spin-isospin interaction is present, the
eigenstates of the Dirac Hamiltonian correspond to combinations
of spin and flavour. If the quark field acts on a fixed flavour
component, it cannot 
create or destroy eigenstates of the chiral
Hamiltonian. 
Therefore the intermediate states $|X\rangle$
obtained {\it e.g.} destroying a quark of specific flavour in the
spin up proton at 
time $t=0$ are not eigenstates of the chiral Hamiltonian and will
evolve in a complicated way in the pionic field, oscillating
in spin and flavour up to the time $t=y_0$.

If the $I^a_{+-}$ function, corresponding to a specific flavour $a$, is indeed
different from zero, it also follows that $I^u_{+-}=-I^d_{+-}$, because the 
usual demonstration can be applied to the flavour averaged quantity,
where one can expand the fields $\psi$ in eigenstates
of the Dirac Hamiltonian. 
Actually this result holds only in first approximation and a more precise
analysis shows that $|I^u_{+-}|$ is not exactly equal to $|I^d_{+-}|$.
This point will be clarified later.

The previous discussion is rather general, and does not refer to any
specific chiral model. It can be applied both to 
sigma-model or to Nambu--Jona--Lasinio model. 
We show now more explicitely how the previous idea is realized
in the sigma-model, whose Lagrangian reads:
\barr
{\cal L} =&i&\bar \psi \gamma^{\mu}\partial_{\mu} \psi
       -g \bar \psi\left(\sigma
+i\gamma_5\vec\tau\cdot\vec\pi\right) \psi\nonumber\\
                       &+&{1\over 2}{\left(\partial_\mu\sigma \right)}^2
                       +{1\over 2}{\left(\partial_\mu\vec\pi\right)}^2
                       -U\left(\sigma ,\vec\pi\right)   \, ,
\earr
where $U(\sigma ,\vec\pi)$ is the usual mexican-hat potential.
In this model the nucleon consists of three valence quarks, moving in a 
background of chiral fields. The pion and the sigma fields are assumed to
be time-independent.

We want to show that: 1) at the single-quark level time reversal
mixes states of different flavour, but 2) the physical nucleon state
obtained from this model satisfies the usual time reversal relation.

To study time-reversal for the single-quark state we write the
Dirac equation. It reads:
$$
[k_\mu \gamma^\mu -g(\sigma+i\gamma_5\vec\tau\cdot\vec\pi)]u(k)=0 \, .$$
We seek now the time reversed solution of the same equation, corresponding
to the substitution $\bfk\rightarrow -\bfk$. Using the standard procedure
one obtains the equation:
$$[k_\mu \gamma^\mu -g(\sigma-i\gamma_5\vec\tau\,^T \cdot\vec\pi)]
\gamma_5 C u^*(\tilde k)=0 \, ,$$
where $\tilde k=(k_0,-\bfk)$ and $C=i\gamma_0 \gamma_2$.  As it can be
seen, if the pion field is absent the time-reversed solution is given
by $\gamma_5 C u^*(\tilde k)$. On the other hand, the term containing
the pion has been modified by the previous transformation.  To
compensate, one needs to introduce an isospin rotation. Since
$(-i\tau_2)(-\vec\tau\,^T)(i\tau_2)=\vec\tau$, the time reversed
solution reads $(-i\tau_2)\gamma_5 C u^*(\tilde k)$ and therefore
under time-reversal the isospin of the single quark is reversed.
Actually, quark states of fixed isospin are not eigenstates of the
chiral hamiltonian.  A good example is provided by the hedgehog which
is used in most practical calculations in chiral models. In the mean
field approximation, taking the hedgehog form for the mean pion field,
$\vec\pi=\hat r \phi(r)$, the spin-isospin wave function of a positive
energy $S$-eigenstate of the chiral hamiltonian is given by
$|h\rangle={1\over \sqrt{2}}[\,|u + \rangle-|d - \rangle]$.

We study now the problem of time reversal for the physical nucleon.
The mean field solution for the three-quark 1$S$-state does not correspond
to a specific spin and flavour state, but it is a superposition of
nucleon and delta.  An eigenstate of spin and flavour can be obtained
using the so-called projection technique, developed by many authors
\cite {many}. They introduce the projector:
\begin{equation}
P_{MM_T}^J\equiv (-1)^{J+M_T} {2J+1\over 8\pi^2}
\int d^3\Omega\,{{\cal D}^{J\textstyle{*}}_{M,-M_T}}(\Omega )R(\Omega ),
\label{310}
\end{equation}
where ${\cal D}^J_{M,K}(\Omega )$ are the
Wigner functions and $R(\Omega )$ is the rotation operator.
Due to the symmetry of the hedgehog (grand-spin zero state), 
this rotation can be performed either on spin or on isospin. 
In the following we shall assume a rotation in spin space.

A baryon state reads:
\begin{equation}
|J=T,J_3=M,T_3=M_T\rangle=P_{MM_T}^J |\psi_h\rangle,
\end{equation}
where $|\psi_h\rangle$ is the mean field solution, made of three
hedgehog quarks surrounded by a coherent state of pions in hedgehog
configuration. Both the quarks and the pions contribute to the spin
and flavour of the proton, because the pions carry orbital angular
momentum $L=1$. This projection technique has been used in many
calculations, including the evaluation of structure functions
\cite{barone}.

It is straightforward to check that the above defined nucleon state
satisfies the usual relation under time reversal. In particular we want
to check that the operator representing time reversal is given by
$${\cal T}=K R_T(\sigma)\, ,$$ 
where $K$ is the complex conjugate operator and $ R_T(\sigma)=-i\sigma_2$.
The projector operator under time reversal transforms as:
\barr
{\cal T} P^J_{MM_T}{\cal T}^\dagger&=&(-1)^{M-M_T}P^J_{-M,-M_T}\nonumber\\
&=&(-1)^{J-M} P^J_{-M,M_T} R_T(\tau)\, ,
\earr
where $ R_T(\tau)=-i\tau_2$.
Since the hedgehog state is invariant under the combined action of
$ R_T(\sigma) R_T(\tau)$, it follows that
$${\cal T}P^J_{MM_T}|\psi_h\rangle=(-1)^{J-M}P^J_{-MM_T}|\psi_h\rangle\, ,$$
which is the usual time-reversal transformation.

Let us now suggest how the $I_{+-}$ function can be computed.
One can proceed as in the computation of the usual distribution
functions, inserting a complete set of states between $\bar\psi$ and $\psi$
and considering only the contributions coming from diquark states, which
should dominate at large $x$. The $I^{u}_{+-}$ function is
dominated by the scalar diquark, while $I^{d}_{+-}$ gets contributions
from the vector diquark only. Since the scalar diquark is lighter than the 
vector one, $I^{u}_{+-}$ extends to larger $x$ than $I^{d}_{+-}$,
in agreement with the results of the phenomenological analysis of 
Ref. \cite{mauro}. It is therefore clear that 
$|I^{u}_{+-}|$ and $|I^{d}_{+-}|$ are not exactly equal.
There is no contradiction with the previous considerations, where it has
been shown that {\it in first approximation} $I^{u}_{+-}=-I^{d}_{+-}$,
because that result holds at mean field level, where there is no difference
between the mass of the scalar and of the vector diquark. 

To conclude, we have shown that quark states of specific spin and flavour 
are not eigenstates of the Hamiltonian. It is indeed possible to build 
states of definite spin and flavour in chiral Lagrangians, but they 
correspond to a mixing of quarks and pions. As a consequence time reversal
invariance does not forbid off-diagonal matrix element of the density 
operator, Eq. (8), to be different from zero, and this can lead to single 
spin asymmetries in $\pup p \to \pi^+ X$ and $\pup p \to \pi^- X$ 
processes, as suggested in Refs. \cite{sivers} and \cite{mauro} 
and in agreement with observation. Such asymmetries would then reveal,
in this interpretation, the presence of an interaction which mixes states with 
different flavours. 

There might be another origin of single spin asymmetries in inclusive 
production of hadrons, related to spin and $\bfk_\perp$ effects in the 
fragmentation of a polarized quark, as suggested in Refs. 
\cite{collins,artcol}; it might then be difficult to single out the specific 
mechanism discussed here if one considers only $\pup p \to \pi X$  processes
or similar ones in which one merely observes a final large $p_T$ hadron. 
However, the possible origin from the quark fragmentation can be excluded by   
looking at single spin asymmetries in $\pup p \to \gamma X$ (no fragmentation)
or even in $\pup p \to \pi X$
or $\ell \pup \to \pi X$, by selecting a final $\pi$, or any other 
hadron, collinear with the jet ($\bfk_\perp = 0$). 

An analysis of single spin asymmetries in DIS has 
recently been performed \cite{alm}, without taking into account the 
spin-isospin interactions considered here: these allow the unique 
possibility of observing single spin asymmetries in fully inclusive DIS 
processes, $\ell \pup \to \ell X$, with unpolarized or longitudinally 
polarized leptons scattering off protons (or neutrons) polarized 
perpendicularly to the scattering plane:

\begin{eqnarray}
&\phantom{=}& \frac{d\sigma^{\ell\pup\to\ell X}}{dx\,dQ^2}-
\frac{d\sigma^{\ell\pdown\to\ell X}}{dx\,dQ^2} \nonumber \\
&=& \sum_{q}\int d\bfk_{\perp}\,
\Delta^{N}f_{q/\pup}(x,\bfk_{\perp})\,
\frac{d\hat{\sigma}^{\ell q\to\ell q}}{dQ^2}(x,\bfk_{\perp})\, .
\label{lplx}
\end{eqnarray}

Our mechanism -- related to the
non vanishing of the off-diagonal matrix elements of the quark density
operator -- would be the only way of obtaining a non zero value for the
single spin asymmetries (\ref{lplx}). Such data might already be
available from polarized DIS experiments and only need a dedicated analysis. 

\acknowledgments

We would like to thank P. Alberto, V. Barone,
L. Caneschi and M. Fiolhais for many useful discussions.

\end{document}